# Advancing Trace Recovery Evaluation – Applied Information Retrieval in a Software Engineering Context

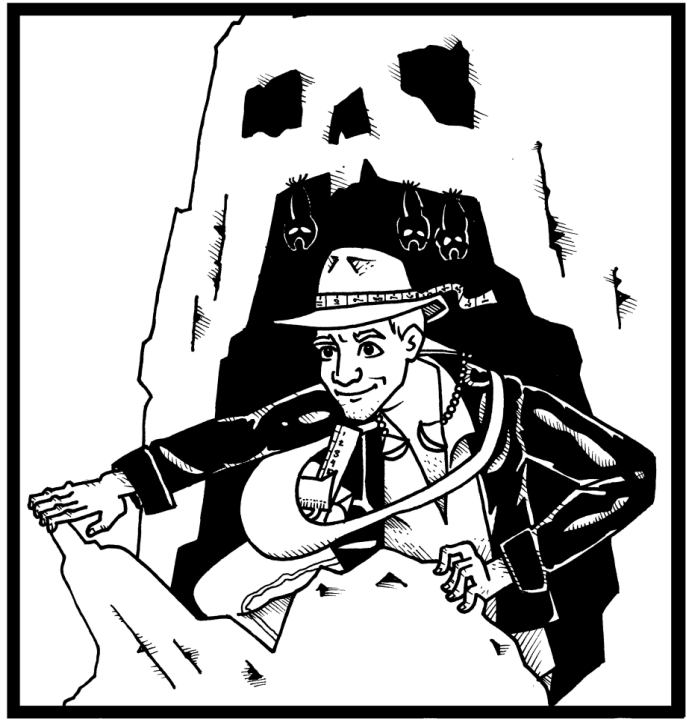

**Markus Borg**

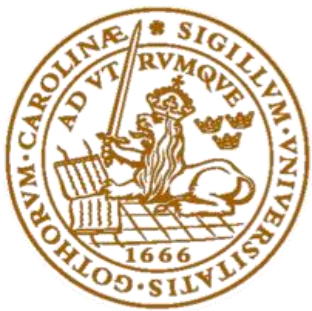

Licentiate Thesis, 2012
Department of Computer Science
Lund University

# ABSTRACT

Successful development of software systems involves the efficient navigation of software artifacts. However, as artifacts are continuously produced and modified, engineers are typically plagued by challenging information landscapes. One state-of-practice approach to structure information is to establish trace links between artifacts; a practice that is also enforced by several development standards. Unfortunately, manually maintaining trace links in an evolving system is a tedious task. To tackle this issue, several researchers have proposed treating the capture and recovery of trace links as an Information Retrieval (IR) problem. The goal of this thesis is to contribute to the evaluation of IR-based trace recovery, both by presenting new empirical results and by suggesting how to increase the strength of evidence in future evaluative studies.

This thesis is based on empirical software engineering research. In a Systematic Literature Review (SLR) we show that a majority of previous evaluations of IR-based trace recovery have been technology-oriented, conducted in "the cave of IR evaluation", using small datasets as experimental input. Also, software artifacts originating from student projects have frequently been used in evaluations. We conducted a survey among traceability researchers and found that while a majority consider student artifacts to be only partly representative of industrial counterparts, such artifacts were typically not validated for industrial representativeness. Our findings call for additional case studies to evaluate IR-based trace recovery within the full complexity of an industrial setting. Thus, we outline future research on IR-based trace recovery in an industrial study on safety-critical impact analysis.

Also, this thesis contributes to the body of empirical evidence of IR-based trace recovery in two experiments with industrial software artifacts. The technology-oriented experiment highlights the clear dependence between datasets and the accuracy of IR-based trace recovery, in line with findings from the SLR. The human-oriented experiment investigates how different quality levels of tool output affect the tracing accuracy of engineers. While the results are not conclusive, there are indications that it is worthwhile further investigating into the actual value of improving tool support for IR-based trace recovery. Finally, we present how tools and methods are evaluated in the general field of IR research, and propose a taxonomy of evaluation contexts tailored for IR-based trace recovery in software engineering.

# ACKNOWLEDGEMENTS

This work was funded by the Industrial Excellence Center EASE – Embedded Applications Software Engineering.

First and foremost, I would like to express my deepest gratitude to my supervisor, Prof. Dr. Per Runeson, for all his support and for showing me another continent. I will never forget the hospitality his family extended to me during my winter in North Carolina.

Secondly, thanks go to my assistant supervisor, Prof. Dr. Björn Regnell, for stimulating discussions in the EASE Theme D project. I am also very grateful to Dr. Dietmar Pfahl. Thank you for giving your time to explain all those things I should have known already. I want to thank my co-authors Krzysztof Wnuk, Dr. Anders Ardö, and Dr. Saïd Assar. All of you have shown me how research can be discussed outside the office sphere. Furthermore, I would like to clearly acknowledge my other colleagues at the Department of Computer Science and the Software Engineering Research Group, as well as my colleagues in the EASE Theme D project and in the SWELL research school.

Moreover, as a software engineer on study leave, I would like to recognize my colleagues at ABB in Malmö. In particular, I want to thank my former manager Henrik Holmqvist for suggesting the position as a PhD student. Also, thanks go to both my current manager Christer Gerding for continual support, and to Johan Gren for being my main entry point regarding communication with the real world.

Finally, I want to thank all my family and friends, and to express my honest gratitude to my parents for always helping me with my homework, and without whose support I wouldn't have begun this journey. And I am indebted to Hannah for the cover art and to Ulla and Ronny for wining and dining me while I wrapped up this thesis. And most importantly, thank you to Marie for all the love and being the most stubborn of reviewers. Of course, you have mattered the most!

*Markus Borg*
*Malmö, August 2012*

Pursuing that doctoral degree
(V) (;,,;) (V)

# LIST OF PUBLICATIONS

## Publications included in the thesis

I **Recovering from a decade: A systematic literature review of information retrieval approaches to software traceability**
*Markus Borg, Per Runeson, and Anders Ardö*
Submitted to a journal, 2012.

II **Industrial comparability of student artifacts in traceability recovery research - An exploratory survey**
*Markus Borg, Krzysztof Wnuk, and Dietmar Pfahl*
In Proceedings of the 16th European Conference on Software Maintenance and Reengineering (CSMR'12), Szeged, Hungary, pp. 181–190, 2012.

III **Evaluation of traceability recovery in context: A taxonomy for information retrieval tools**
*Markus Borg, Per Runeson, and Lina Brodén*
In Proceedings of the 16th International Conference on Evaluation & Assessment in Software Engineering (EASE'12), Ciudad Real, Spain, pp. 111–120, 2012.

IV **Do better IR tools improve software engineers' traceability recovery?**
*Markus Borg, and Dietmar Pfahl*
In Proceedings of the International Workshop on Machine Learning Technologies in Software Engineering (MALETS'11), Lawrence, KS, USA, pp. 27–34, 2011.

The following papers are related, but not included in this thesis. The research agenda presented in Section 8 is partly based on Paper V, which was published as a position paper. Paper VI contains a high-level discussion on information management in a large-scale software engineering context.

# Related publications

V **Findability through traceability - A realistic application of candidate trace links?**
*Markus Borg*
In Proceedings of 7th International Conference on Evaluation of Novel Approaches to Software Engineering (ENASE'12), Wrocław, Poland, pp. 173–181, 2012.

VI **Towards scalable information modeling of requirements architectures**
*Krzysztof Wnuk, Markus Borg, and Saïd Assar*
To appear in the Proceedings of the 1st International Workshop on Modelling for Data-Intensive Computing, (MoDIC'12), Florence, Italy, 2012.

# Contribution statement

Markus Borg is the first author of all included papers. He was the main inventor and designer of the studies, and was responsible for running the research processes. Also, he conducted most of the writing.

The SLR reported in Paper I was a prolonged study, which was conducted in parallel to the rest of the work included in this licentiate thesis. The study was co-designed with Prof. Per Runeson, but Markus Borg wrote a clear majority of the paper. Furthermore, Mats Skogholm, a librarian at the Lund University library contributed to the development of search strings, and Dr. Anders Ardö validated the data extraction process and reviewed technical contents of the final report.

The survey in Paper II was conducted by Markus Borg and Krzysztof Wnuk. They co-designed the study and distributed the questionnaire in parallel, however Markus Borg was responsible for the collection and analysis of the data. Markus Borg wrote a majority of the report, however with committed assistance from Dr. Dietmar Pfahl and Krzysztof Wnuk.

The experimental parts of Paper III started as a master thesis project by Lina Brodén. Markus Borg was the main supervisor, responsible for research design and collection of data from industry. Prof. Per Runeson had the role as examiner, and also gave initial input to the study design. The outcome of the master thesis project was then used by Markus Borg as input to Paper III, who extended it with a discussion on evaluation contexts.

The experimental setup in Paper IV was originally designed by Markus Borg, and then further improved together with Dr. Dietmar Pfahl. Markus Borg executed the experiment, analyzed the data, and conducted most of the writing. Dr. Dietmar Pfahl also contributed as an active reviewer and initiated rewarding discussions.

# CONTENTS

# 1 Introduction

Modern society depends on software-intensive systems. Software operates invisibly in everything from kitchen appliances to critical infrastructure, and living a life without daily relying on systems running software requires a determined downshifting from life as most people enjoy it. As the significance of software continuously grows, so does the importance of being able to create it efficiently.

*Software development* is an inclusive expression used to describe any approach to produce source code and its related documentation. During the software crisis of the 1960s, it became clear that software complexity quickly rises when scaled up to larger systems. The development methods that were applied at the time did not result in required software in a predictable manner. *Software engineering* was coined to denote software developed according to a systematic and organized approach, aiming to effectively produce high-quality software with reduced uncertainty [72]. By applying the engineering paradigm to software development, activities such as analysis, specification, design, implementation, verification, and evolution turned into well-defined practices. On the other hand, additional knowledge-intensive activities tend to increase the number of documents maintained in a project [97].

Large projects risk being characterized by *information overload*, a state where individuals do not have time or capacity to process all available information [33]. Knowledge workers frequently report the stressing feeling of having to deal with too much information [32], and in general spend a substantial effort on locating relevant information [59, 65]. Also, in software engineering the challenge of dealing with a plentitude of software artifacts has been highlighted [37, 74]. Thus, an important characteristic of a software engineering organization is the *findability* it provides, herein defined as "the degree to which a system or environment supports navigation and retrieval" [71], particularly in globally distributed development [28].

One state-of-practice way to support findability in software engineering is to maintain *trace links* between artifacts. *Traceability* is widely recognized as an important factor for efficient software engineering [5, 18, 29]. Traceability supports engineering activities related to software evolution, e.g., change management, impact analyses, and regression testing [5, 16]. Also, traceability assists engineers in less concrete tasks such as system comprehension, knowledge transfer, and process alignment [23, 80, 85]. On the other hand, maintaining trace links in an evolving system is known to be a tedious task [29, 34, 48]. Thus, to support trace link maintenance, several researchers have proposed tool support for *trace recovery*, i.e., proposing candidate trace links among existing artifacts, based on *Information Retrieval* (IR) approaches [5, 23, 48, 68, 70]. The rationale is that IR refers to a set of techniques for finding relevant documents from within large collections [6, 69], and that the search for trace links can be interpreted as an attempt to satisfy an information need.

This thesis includes an aggregation of empirical evidence of *IR-based trace*

*recovery*, and contributes to the body of knowledge on conducting evaluative studies on IR-based trace recovery tools. As applied researchers, our main interest lies in understanding how feasible trace recovery tools would be for engineers in industrial settings. Paper I contains a comprehensive literature review of previous research on the topic. Based on questions that arose during the literature review, other studies were designed and conducted in parallel. Paper II provides an increased understanding of the validity of using artifacts originating from student projects as experimental input. Paper III reports from an experiment with trace recovery tools on artifacts collected from industry, and also proposes a taxonomy of evaluation contexts tailored for IR-based trace recovery. Finally, Paper IV presents a novel experiment design, addressing the value of slight tool improvements.

This introduction chapter provides a background for the papers and describes relationships between studies. The remainder of this chapter is organized as follows. Section 2 presents a brief background of traceability research and introduces fundamentals of IR. Also, it presents how IR can be applied to address trace recovery. Section 3 expresses the overall aim of this thesis as research questions, and proposes three viewpoints from which the results can be interpreted. Also, Section 4 presents the research methodologies used to answer the research questions. The results of each individual paper are presented in Section 5, while Section 6 draws together the results to provide answers to the research questions. Section 7 highlights some threats to validity in the presented research. In Section 8, we present how we plan to continue our work in future studies. Finally, Section 9 concludes the introduction of this thesis.

## 2  Background and Related Work

This section presents a background of traceability and IR, the two main research areas on which this thesis rests. Also, we present how IR has been applied to support trace link maintenance. Finally, we present related work on advancing IR-based trace recovery evaluation.

### 2.1  Traceability - A Fundamental Software Engineering Challenge

The concept of traceability has been discussed in software engineering since the very beginning. Already at the pioneering NATO Working Conference on Software Engineering in 1968, a paper by Randall recognized the need for a developed software system to "contain explicit traces of the design process" [81]. In an early survey of software engineering state-of-the-art in the 1970s by Boehm, "traceability" was mentioned six times, both in relation to engineering challenges at the time, and when predicting future trends [12]. In the software industry, traceability became acknowledged as an important aspect in high quality development.

Consequently, by the 1980s several development standards specified process requirements on the maintenance of traceability information [30]. At the time, the IEEE definition of traceability was "the degree to which a relationship can be established between two or more products of the development process, especially products having a predecessor-successor or master-subordinate relationship to one another; for example, the degree to which the requirements and design of a given software component match" [50].

In the 1990s, the requirements engineering community emerged and established dedicated publication fora. As traceability was in scope, published research on the topic increased and its relation to requirements engineering was further emphasized. A paper by Gotel and Finkelstein in 1994 identified the lack of a common definition of requirements traceability and suggested: "Requirements traceability refers to the ability to describe and follow the life of a requirement, in both a forwards and backwards direction (i.e., from its origins, through its development and specification, to its subsequent deployment and use, and through all periods of on-going refinement and iteration in any of these phases)" [40]. According to a recent systematic literature review by Torkar *et al.*, this definition of requirements traceability is the most commonly cited in research publications [91].

While traceability has been questioned by some of the lean-thinkers of the agile movement in the 2000s to be too costly in relation to its benefits [79], traceability continues to be a fundamental aspect in many development contexts. Since traceability is important to software verification, general safety standards such as IEC 61508 [53], and industry-specific variants (e.g., ISO 26262 in the automotive industry [54] and IEC 61511 in the process industry [52]) mandate maintenance of traceability information. Furthermore, as traceability has a connection to quality, it is also required by organizations aiming at process improvement as defined by CMMI (Capability Maturity Model Integration) [14]. Thus, traceability is neither negotiable in safety certifications nor in CMMI appraisals. In 2006, the international organization CoEST, the Center of Excellence for Software Traceability[1], was founded to gather academics and practitioners to advance traceability research.

In the beginning of 2012, an extensive publication edited by Cleland-Huang *et al.* was published [18], containing contributions from several leading researchers in the traceability community. Apart from summarizing various research topics on traceability, the work presents a number of fundamental definitions. We follow the proposed terminology in the introduction chapter of this thesis. However, when the included Papers II-IV were written, the terminology was not yet aligned in the community. Cleland-Huang *et al.* define traceability as: "the potential for traces to be established and used" [39], i.e., the trace *ability* is stressed. On the other hand, they also present the more specific definition of *requirements traceability* by Gotel and Finkelstein in its original form.

---

[1] www.coest.org

A number of other terms relevant for this thesis are also defined in the book. A *trace artifact* denotes a traceable unit of data. A *trace link* is an association forged between two trace artifacts, representing relations such as overlap, dependency, contribution, evolution, refinement, or conflict [80]. The main subject in this thesis, *trace recovery*, denotes an approach to create trace links among existing software artifacts. This is equivalent to what we consistently referred to as *traceability recovery* in the Papers II-IV.

## 2.2 Information Retrieval - Satisfying an Information Need

A central concept in this thesis is *information seeking*, the "conscious effort to acquire information in response to a gap in knowledge" [15]. Particularly, we are interested in finding pieces of information that enable trace links to be recovered, i.e. *trace link seeking*. One approach to seek information is *information retrieval*, meaning "finding material (usually documents) of an unstructured nature (usually text) that satisfies an information need from within large collections (usually stored on computers)" [69]. This section briefly presents the two main categories of IR models. A longer presentation is available in Paper I.

Typically, IR models apply the bag-of-words model, a simplifying assumption that represents a document as an unordered collection of words, disregarding word order [69]. Most IR models can be classified as either *algebraic* or *probabilistic*, depending on how relevance between queries and documents is measured. Algebraic IR models assume that relevance is correlated with similarity, while probabilistic retrieval is based on models estimating the likelihood of queries being related to documents.

The *Vector Space Model* (VSM), developed in the 1960s, is the most commonly applied algebraic IR model [86]. VSM represents both documents and queries as vectors in a high-dimensional space and similarities are calculated between vectors using distance functions. In principle, every term constitutes a dimension. Usually, terms are weighted using some variant of Term Frequency-Inverse Document Frequency (TF-IDF). TF-IDF is used to weight a term based on the length of the document and the frequency of the term, both in the document and in the entire document collection. *Latent Semantic Indexing* (LSI) is an approach to reduce the dimension space, sometimes successful in reducing the effects of synonymy and polysemy [25]. LSI uses singular value decomposition to transform the dimensions from individual terms to combinations of terms, i.e., concepts constitute the dimensions rather than individual terms.

In probabilistic models, documents are ranked according to their probability of being relevant to the query. Two common models are the *Binary Independence retrieval Model* (BIM) [25] [82] and *Probabilistic Inference Networks* [92]. A subset of probabilistic retrieval estimate *Language Models* (LM) for each document. Documents are then ranked based on the probability that a document would generate the terms of a query [78]. A later refinement of simple LMs, topic models, describes documents as a mixture over topics, where each topic is characterized by an LM. Examples include probabilistic latent semantic indexing [45] and Latent Dirichlet Allocation (LDA) [11].

## 2.3 Information Retrieval Evaluation

IR is a highly experimental discipline, and empirical evaluations are the main research tool to scientifically compare IR algorithms. The state-of-the-art has advanced through careful examination and interpretation of experimental results. Traditional IR evaluation, as it was developed by Cleverdon in the Cranfield project in the late 1950s [20], consists of three main elements: a document collection, a set of information needs (typically formulated as queries), and relevance judgments telling what documents are relevant to these information needs, i.e., a *gold standard*. Thus, as the experimental units are central in IR evaluation it is important to address threats to content validity, i.e., the extent to which the experimental units reflect and represent the elements of the domain under study [93]. A selection of experimental units that match the real-world setting should carefully be selected, and the sample should be sufficiently large to be representative to the domain.

The most common way to evaluate the effectiveness of an IR system is to measure *precision* and *recall*. As displayed in Figure 1, precision is the fraction of retrieved documents that are relevant, while recall is the fraction of relevant documents that are retrieved. The outcome is often visualized as a Precision-Recall (P-R) curve where the average precision is plotted at fixed recall values, presented as PR@Fix in Figure 1. However, this set-based approach has been criticized for being opaque, as the resulting curve obscure the actual numbers of retrieved documents needed to get beyond low recall [90]. Alternatively, the ranking of retrieval results can be taken into account. The most straightforward approach is to plot the P-R curve for the top k retrieved documents instead [69], shown as PR@k in Figure 1. In such curves, one can see the average accuracy of the first search result, the first ten search results etc. The Text Retrieval Conference (TREC), hosting the most distinguished evaluation series for IR, reports results using both precision at 11 fixed recall values (0.0, 0.1 ... 1.0) and precision at the top 5, 10, 15, 30, 100 and 200 retrieved documents [90]. A discussion on IR-based trace recovery evaluation styles is available in Paper I.

There are several other measures available for IR evaluations, including F-measure, ROC curve, R-precision and the break-even point [69], but none of them are as established as P-R curves. On the other hand, two measures offering single figure effectiveness have gained increased attention. *Mean Average Precision* (MAP), roughly the area under the P-R curve for a set of queries, is established in the TREC community. *Normalized Discounted Cumulative Gain* [56], similar to precision at top k retrieved documents but especially designed for non-binary relevance judgments, is popular especially among researchers employing machine learning techniques to rank search results.

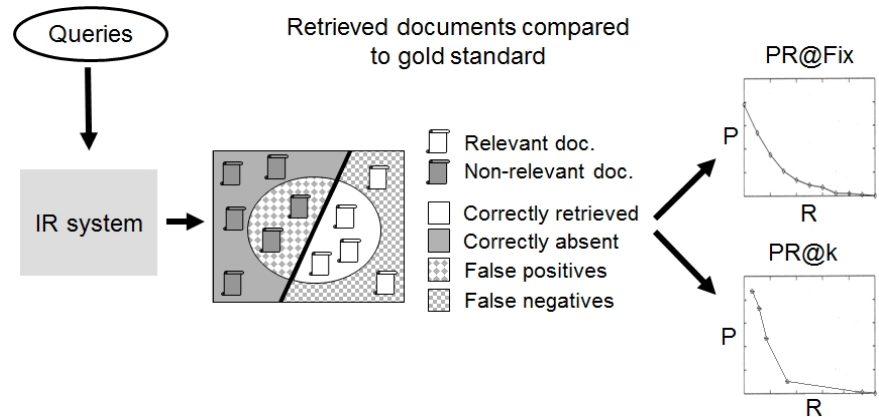

**Figure 1:** Traditional IR evaluation using P-R curves showing PR@Fix and PR@k. In the center part of the figure, displaying a document space, the relevant items are to the right of the straight line while the retrieved items are within the oval.

The experimental setups of IR evaluations rarely fulfil assumptions required for significance testing, e.g., independence between retrieval results, randomly sampled document collections, and normal distributions. Thus, traditional statistics has not had a large impact on IR evaluation [41]. However, it has been proposed both to use hypergeometric distributions to compute retrieval accuracy to be expected by chance [87], and to apply the Monte Carlo method [13].

Although IR evaluation has been dominated by technology-oriented experiments, it has also been challenged for its unrealistic lack of user involvement [61]. Ingwersen and Järvelin argued that IR is always evaluated in a context and proposed an evaluation framework, where the most simplistic evaluation context is referred to as "the cave of IR evaluation" [51]. In Paper III, we present their framework and an adapted version tailored for IR-based trace recovery.

## 2.4 Trace Recovery - An Information Retrieval Problem

Tool support for linking artifacts containing Natural Language (NL) has been explored by researchers since at least the early 1990s. Whilst a longer history of IR-based trace recovery is available in Paper I, this section introduces the approach and highlights some key publications.

The underlying assumption of using IR for trace recovery is that artifacts with highly similar textual content are good candidates to be linked. Figure 2 shows the key steps involved in IR-based trace recovery, organized in a pipeline architecture as suggested by De Lucia *et al.* [24]. First, input documents are parsed and preprocessed, typically using *stop word removal* and *stemming*. In the second step, the documents are indexed using the IR model. Then, candidate trace links are generated, ranked according to the IR model, and the result is visualized. Finally, the result is presented to the engineer, as emphasized in Figure 2, who gets to assess the output. The rationale is that it is faster for an engineer to assess a ranked list of candidate trace links, despite both missed links and false positives, than seeking trace links from scratch.

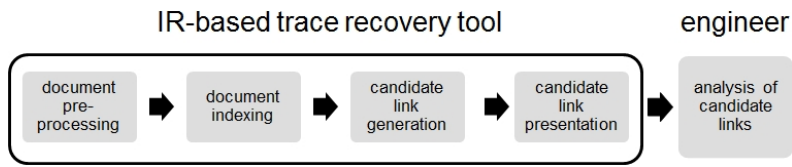

**Figure 2:** Key steps in IR-based trace recovery, adapted from De Lucia *et al.* [24].

In 1998, a pioneering study by Fiutem and Antoniol proposed a trace recovery process to bridge the gap between design and code, based on edit distances between NL content of artifacts [36]. They coined the term "traceability recovery", and published several papers on the topic. Also, they were the first to express identification of trace links as an IR problem [4]. Their well-cited work from 2002 compared the accuracy of candidate trace links from two IR models, BIM and VSM [5]. Marcus and Maletic were the first to apply LSI to recover trace links between source code and NL documentation [70]. Huffman Hayes *et al.* enhanced trace recovery based on VSM with relevance feedback. They had from early on a human-oriented perspective, aiming at supporting V&V activities at NASA using their tool RETRO [49]. De Lucia *et al.* have conducted work focused on empirically evaluating LSI-based traceability recovery in their document management system ADAMS [22]. They have advanced the empirical foundation by conducting a series of controlled experiments and case studies with student subjects. Cleland-Huang and colleagues have published several studies using probabilistic IR models for trace recovery, implemented in their tool Poirot [66]. Much of their work has focused on improving the accuracy of their tool by various enhancements.

Lately, a number of publications suggest that the P-R differences for trace recovery between different IR models are minor. Oliveto *et al.* compared VSM, LSI, LM and LDA on artifacts originating from two student projects, and found no significant differences [73]. Also a review by Binkley and Lawrie reported the same phenomenon [10]. Falessi *et al.* proposed a taxonomy of algebraic IR models and experimentally studied how differently configured algebraic IR models performed in detecting duplicated requirements [34]. They concluded that simple IR solutions tend to produce more accurate output.

We have identified some progress related to IR-based trace recovery in non-academic environments. In May 2012, a US patent with the title "System and method for maintaining requirements traceability" was granted [9]. The patent ap-plication, filed in 2007, describes an application used to synchronize artifacts in a requirements repository and a testing repository. Though the actual linking process is described in general terms, a research publication implementing trace recovery based on LSI is cited. Also indicating industrial interest in IR-based trace recovery is that HP Quality Center, a component of HP Application Lifecycle Management, provides a feature to link artifacts based on textual similarity analysis [44]. While it is implemented to detect duplicate defect reports, the same technique could be applied for trace recovery.

## 2.5 Previous Work on Advancing IR-based Trace Recovery Evaluation

While there are several general guidelines on software engineering research (e.g., experiments [95], case studies [84], reporting [57], replications [88], literature reviews [62]), only few publications have specifically addressed research on IR-based trace recovery. The closest related work is described in this section.

Huffman Hayes and Dekhtyar proposed a framework for comparing experiments on requirements tracing [47]. Their framework describes the four experimental phases: definition, planning, realization and interpretation. We evaluated the framework in Paper III, and concluded that it provides valuable structure for conducting technology-oriented experiments. However, concerning human-oriented experiments, there is room for enhancements. Huffman Hayes *et al.* also suggested categorizing trace recovery research as either *studies of methods* (are the tools capable of providing accurate results fast?) or *studies of human analysts* (how do humans use the tool output?) [48]. Furthermore, in the same publication, they suggested assessing the accuracy of tool output according to the quality intervals "Acceptable/Good/Excellent", with specified P-R levels. In Paper I, we catalog the primary publications according to their suggestions, but we also catalog the primary publications according to the context taxonomy we propose in Paper III. A recent publication by Falessi *et al.* proposed seven "empirical principles" for technology-oriented evaluation of IR tools in software engineering, explicitly mentioning trace recovery as one application [35]. Despite the absence of statistical analysis in traditional IR evaluation [41], they argued for both increased difference and equivalence testing. In Paper IV, we also propose equivalence testing of IR-based trace recovery, however in the context of human-oriented experiments.

A number of researchers connected to CoEST have repeatedly argued that a repository of benchmarks for trace recovery research should be established, in line with what has driven large scale IR evaluations at TREC [17, 26, 27]. Furthermore, Ben Charrada *et al.* have presented a possible benchmark for traceability studies, originating from an example in a textbook on software design [8]. We support the attempt to develop large public datasets, but there are also risks involved in benchmarks. As mentioned in Section 2.3, the results of IR evaluations depend on the experimental input. Thus, there is a risk of over-engineering tools on datasets that do not have acceptable content validity. Benchmarking in IR-based trace recovery is further discussed in Paper III.

| **Tool developer** (Paper I, Paper IV) | **Which IR model should we implement in our new trace recovery tool?** |
|---|---|
| RQ1 | Which IR model has most frequently been implemented in research tools? |
| RQ2 | Which IR model has displayed the most promising results? |
| **Development manager** (Paper I, II) | **Should we deploy an IR-based trace recovery tool in our organization?** |
| RQ3 | What evidence is there that IR-based trace recovery is feasible in an industrial setting? |
| **Traceability researcher** (Paper I, III) | **How can we strengthen the base of empirical evidence of IR-based trace recovery?** |
| RQ4 | How can we advance technology-oriented studies on IR-based trace recovery? |
| RQ5 | How can we advance human-oriented studies on IR-based trace recovery? |

**Table 1:** Viewpoints further broken down into RQs. The scope of trace recovery is implicit in RQ1-2, while explicit in RQ3-5.

Another project, also promoted by CoEST, is the TraceLab project [17, 60]. TraceLab is a visual experimental workbench, highly customized for traceability research. It is currently under Alpha release to project collaborators. CoEST claims that it can be used for designing, constructing, and executing trace recovery experiments, and facilitating evaluation of different techniques. TraceLab is said to be similar to existing data mining tools such as Weka and RapidMiner, and aims at providing analogous infrastructure to simplify experimentation on traceability. However, to what extent traceability researchers are interested in a common experimental workbench remains an open question.

## 3 Research Focus

This section describes how this thesis contributes to the body of knowledge on IR-based trace recovery in general, and evaluations of IR-based trace recovery in particular. We base the discussion on the viewpoints from the perspectives of three stakeholders, each with his own pictured consideration: (1) a CASE tool developer responsible for a new trace recovery tool, (2) a manager responsible for a large software development organization, and (3) an academic researcher trying to advance the trace recovery research frontier. We further divide each viewpoint into more specific Research Questions (RQ), as presented in Table 1.

As the included papers are closely related, all papers to some extent contribute to the three viewpoints. Paper I is the most recent publication included in this thesis, and by far the most comprehensive. As such, it contributes to all individual RQs presented in Table 1. Papers II, III, and IV primarily address the RQs from the perspective of a development manager, an academic researcher, and a tool developer respectively. Figure 3 positions this thesis in relation to existing research on IR and its application on traceability. The rows in the figure show three different research foci: development and application of retrieval models, improving technology-oriented IR evaluations, and improving human-oriented IR evaluations. The arrows A-C indicate how approaches from the IR domain have been applied in traceability research.

- The A arrow represents pioneering work on applying IR models to trace recovery, such as presented by Antoniol *et al.* [5] and Marcus and Maletic [70].

- The B arrow denotes contributions to technology-oriented evaluations of IR-based trace recovery, inspired by methods used in the general IR domain. Examples include the experimental framework proposed by Huffman Hayes and Dekhtyar [47] and CoEST's ambition to create benchmarks [17].

- The context taxonomy we propose in Paper III is an example of work along arrow C, where we apply results from human-oriented IR evaluations to traceability research.

To the right in Figure 3, the internal relations among the included papers are presented. The study in Paper II on industrial comparability of student artifacts was initiated to further explore early results from the work in Paper I. The experiment on human subjects in Paper IV was designed to further analyze the differences between the technology-oriented experimental results in Paper III. Finally, the evaluation taxonomy proposed in paper IV was used in Paper I to structure parts of the literature review.

The right box in Figure 3 also shows to which research focus the included papers mainly contribute. Paper I contributes to all foci. Paper II mainly contributes to improving technology-oriented IR evaluations, as it addresses the content validity of experimental input originating from student projects. Paper III proposes a taxonomy of evaluation contexts, and thus contributes to improving human-oriented IR evaluations. Finally, while Paper IV questions the implications of minor differences of tool output in previous technology-oriented experiments, we consider it to mainly contribute by providing empirical results on the application of IR models on a realistic work task involving trace recovery.

## 4  Method

The research in this thesis is mainly based on *empirical research*, a way to obtain knowledge through observing and measuring phenomena. Empirical studies result in evidence, pieces of information that support conclusions. Since evidence is needed to build and verify theories [63], empirical studies should be conducted to improve the body of software engineering knowledge [31, 84, 89]. However, as engineers, we also have an ambition to create innovations that advance the field of software engineering. One research paradigm that seeks to create innovative artifacts to improve the state-of-practice is *design science* [43]. Design science originates from engineering and is fundamentally a problem solving paradigm. As presented in Figure 4, the build-evaluate loop is central in design science (in some disciplines referred to as action research [83]). Based on empirical understanding of a context, innovations in the form of tools and practices are developed. The innovations are then evaluated empirically in the target context. These steps might then be iterated until satisfactory results have been reached.

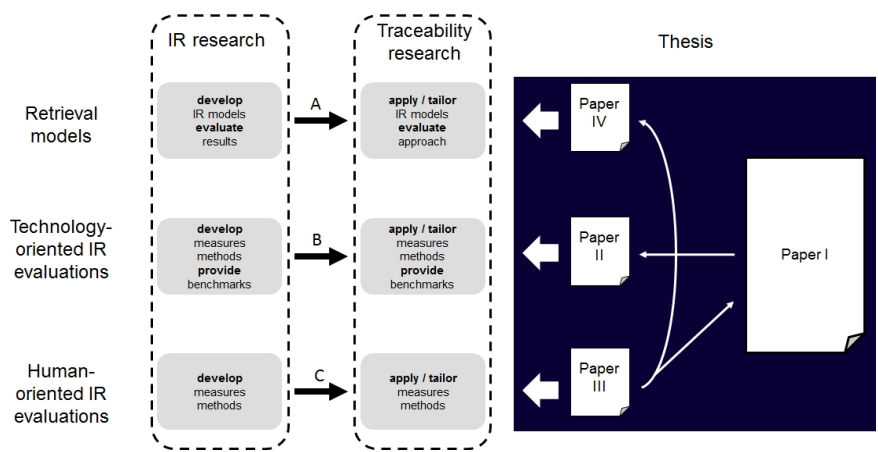

**Figure 3:** Contributions of this thesis, presented in relation to research on IR and traceability. Arrows A-C denote examples of knowledge transfer between the domains. The right side of the figure positions the individual papers of this thesis, and shows their internal relationships.

This thesis mainly contains *exploratory* and *evaluative* empirical research, based on studies using *experiments*, *surveys*, and *systematic literature reviews* as research methodology. However, also *case studies* are relevant for this thesis, as such studies are discussed as possible future work in Section 8. Future work also involves *design tasks* to improve industry practice, after proper assessment.

*Exploratory research* is typically conducted in early stages of research projects, and attempts to bring initial understanding to a phenomenon, preferably from rich qualitative data [31]. An exploratory study is seldom used to draw definitive conclusions, but is rather used to guide further work. Decisions on future study design, data collection methods, and sample selections can be supported by preceding exploratory research. Paper I and Paper II explored both tool support for IR-based

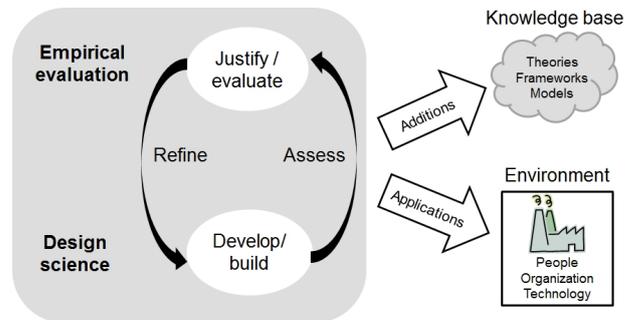

**Figure 4:** Design science and the build-evaluate loop. Adapted from Hevner *et al.* [43]

trace recovery in general, and evaluation of such tools in particular.

*Evaluative research* can be conducted to assess the effects and effectiveness of innovations, interventions, practices etc. [83]. An evaluative study involves a systematic collection of data, which can be of both qualitative and quantitative nature. Paper III contains a quantitative evaluation of two IR-based trace recovery tools compared to a naïve benchmark. Paper IV reports from an evaluation comparing how human subjects solve a realistic work task when supported by IR-based trace recovery. Both papers use industrial artifacts as input to the evaluative studies.

## 4.1 Research Methods

*Experiments* (or controlled experiments) are commonly used in software engineering to investigate the cause-effect relationships of introducing new methods, techniques or tools. Different treatments are applied to or by different subjects, while other variables are kept constant, and the effects on outcome variables are measured [95]. Experiments are categorized as either *technology-oriented* or *human-oriented*, depending on whether objects or human subjects are given various treatments. Involving human subjects is expensive, consequently university students are commonly used and not engineers working in industry [46]. Paper III presents a technology-oriented experiment, while Paper IV describes an experimental setup of a human-oriented experiment, and results from a pilot run using both students and senior researchers as subjects.

A *case study* in software engineering is conducted to study a phenomenon within its real-life context. Such a study draws on multiple sources of evidence to investigate one or more instances of the phenomenon, and is especially applicable when the boundary between phenomenon and its context cannot be clearly specified [84]. While this thesis does not include any case studies, such studies are planned as future work, and are further discussed in Section 8.

| Works | Type of research | Research method |
|---|---|---|
| Paper I | Exploratory | Systematic literature review |
| Paper II | Exploratory | Questionnaire-based survey |
| Paper III | Evaluative | Technology-oriented experiment |
| Paper IV | Evaluative | Human-oriented experiment |

**Table 2:** Type and method of research in the included papers.

A *Systematic Literature Review* (SLR) is a secondary study aimed at aggregating a base of empirical evidence. It is an increasingly popular method in software engineering research, relying on a rigid search and analysis strategy to ensure a comprehensive collection of publications [62]. A variant of an SLR is a Systematic Mapping (SM) study, a less granular literature study, designed to identify research gaps and direct future research [62, 76]. The primary contribution of this thesis comes from the SLR presented in Paper I.

*Survey* research is used to describe what exists, and to what extent, in a given population [55]. Three distinctive characteristics of survey research can be identified [77]. First, it is used to quantitatively (sometimes also qualitatively) describe aspects of a given population. Second, the collected data is collected from people and thus subjective. Third, survey research uses findings from a portion of a population to generalize back to the entire population. Surveys can be divided into two categories based on how they are executed: written surveys (i.e., questionnaires) and verbal surveys (i.e., telephone or face-to-face interviews). Paper II reports how valid traceability researchers consider studies on student artifacts to be, based on empirical data collected using a questionnaire-based survey.

Table 2 summarizes the type of research, and the selected research method, in the included papers.

## 5 Results

This section presents the main results from each of the included papers.

### Paper I: A Systematic Literature Review of IR-based Trace Recovery

The objective of the study was to conduct a comprehensive review of IR-based trace recovery. Particularly focusing on previous evaluations, we explored collected evidence of the feasibility of deploying an IR-based trace recovery tool in an industrial setting. Using a rigorous methodology, we aggregated empirical data from 132 studies reported in 79 publications. We found that a majority of the publications implemented algebraic IR models, most often the classic VSM. Also, we found that no IR model regularly outperforms VSM. Most studies do not analyze the usefulness of the IR-based trace recovery further than tool output, i.e., evaluations conducted "in the cave", entirely based on P-R curves dominate. The strongest evidence of the benefits of IR-based trace recovery comes from a number of controlled experiments. Several experiments report that subjects perform certain software engineering work tasks faster (and/or with higher quality) when supported by IR-based trace recovery tools. However, the experimental settings have been artificial using mainly student subjects and small sets of software artifacts. In technology-oriented evaluations, we found a clear dependence between datasets used in evaluations and the experimental results. Also, we found that few case studies have been conducted, and only one in an industrial context. Finally, we conclude that the overall quality of reporting should be improved regarding both context and tool details, measures reported, and use of IR terminology.

## Paper II: Researchers' Perspectives on the Validity of Student Artifacts

While conducting the SLR in Paper I, we found that in roughly half of the evaluative studies on IR-based trace recovery, output from student projects was used as experimental input. Paper II explores to what extent student artifacts differ from industrial counterparts when used in evaluations of IR-based trace recovery. In a survey among authors identified in the SLR in Paper I, including both academics and practitioners, we found that a majority of the respondents consider software artifacts originating from student projects to be only partly comparable to industrial artifacts. Moreover, only few respondents reported that they validated student artifacts for industrial representativeness before using them as experimental input. Also, our respondents made suggestions for improving the description of artifact sets used in IR-based trace recovery studies.

## Paper III: A Taxonomy of Evaluation Contexts and a Cave Study

Paper III contains a technology-oriented experiment, an evaluation "in the cave", of two publicly available IR-based trace recovery tools. We use both a de-facto traceability benchmark originating from a NASA project, and artifacts collected from a company in the domain of process automation. Our study shows that even though both artifact sets contain software artifacts from embedded development, their characteristics differ considerably, and consequently the accuracy of the recovered trace links. Furthermore, Paper III proposes a context taxonomy for evaluations of IR-based trace recovery, covering evaluation contexts from "the cave" to in-vivo evaluations in industrial projects. This taxonomy was then used to structure parts of the SLR in Paper I.

**Paper IV: Towards Understanding Minor Tool Improvements**

Since a majority of previous studies evaluated IR-based trace recovery only based on P-R curves, one might wonder to what extent minor improvements of tool output actually influence engineers working with the tools. Is it worthwhile to keep hunting slight improvements in precision and recall? To tackle this question, we conducted a pilot experiment with eight subjects, supported by tool output from the two tools evaluated in Paper IV. As such, the subjects were supported by tool output matching two different P-R curves. Inspired by research in medicine, more specifically a study on vaccination coverage [7], we then analyzed the data using statistical testing of equivalence [94]. The low number of subjects did not result in any statistically significant results, but we found that the effect size of being supported by the slightly more accurate tool output was of practical significance. While our results are not conclusive, the pilot experiment indicates that it is worthwhile to investigate further into the actual value of improving tool support for trace recovery, in a replication with more subjects.

# 6 Synthesis

This section presents a synthesis of the results from the included papers to provide answers to the research questions asked in this thesis.

## RQ1: Which IR model has most frequently been implemented in research tools?

Paper I concludes that algebraic IR models have been implemented more often than probabilistic IR models. Although there has been an increasing trend of trace recovery based on probabilistic LMs the last five years, a majority of publications report IR-based trace recovery using vector space retrieval. In roughly half of the papers applying VSM, the number of dimensions of the vector space is reduced using LSI. However, it is important to note that one reason for the many studies on vector space retrieval is that it is frequently used as a benchmark when comparing the output from more advanced IR models.

## RQ2: Which IR model has displayed the most promising results?

As presented in Paper I, no IR model has been reported to repeatedly outperform the classic VSM developed in the 60s. This confirms previous work by Oliveto *et al.* [73], Binkley and Lawrie [10], and Falessi *et al.* [34]. Instead, our work shows that the input software artifacts have a much larger impact on the outcome of trace recovery experiments than the choice of IR model. While this is well known in IR research [69], and has been mentioned by Ali *et al.* in the traceability community [2], it has not been highlighted as clearly before in traceability research.

### RQ3: What evidence is there that IR-based trace recovery is feasible in an industrial setting?

The software engineering literature identified in Paper I does not contain any substantial success stories from in-vivo evaluations. Only one industrial case study, conducted in a short 5-people project, has reported that IR-based trace recovery was beneficial. Apart from this study, the strongest empirical evidence comes from controlled experiments with student subjects (similar to our contribution in Paper IV) and case studies in student projects. While these studies suggest that certain traceability-centric work tasks can be supported by IR-based trace recovery tools, the majority of studies do not go further than reporting P-R curves "in the cave of IR evaluation" (similar to our contribution in Paper III). However, some identified non-academic activity indicates a usefulness of the approach. In May 2012, a patent was granted protecting a "System and method for maintaining requirements traceability" [9]. Furthermore, the CASE tool HP Quality Center describes an IR feature in its marketing of the product [44].

### RQ4: How can we advance technology-oriented studies on IR-based trace recovery?

As the results of IR-based trace recovery are so dependent on the input software artifacts, there is little value in additional evaluations based on a small number of artifacts. It is critical to conduct experiments on large, preferably publicly available, datasets. While this has been proposed by members of COEST before [8, 17, 26, 27], Paper I underlines how few previous evaluations have been conducted using datasets of reasonable size. Moreover, in Paper II we argue that if student artifacts are to be used as experimental input, they should first be properly validated for industrial representability. In the IR sub-domain of enterprise search, it has been proposed to extract documents from companies that no longer exist [42] (e.g., Enron), an option that could be explored also in software engineering. In Paper I we argue that the reporting of technical details of IR implementations should be improved, while Paper II stresses the importance to clearly describe the input artifacts in technology-oriented experiments. Describing the artifacts is especially important in studies where the artifacts cannot be disclosed, e.g., for confidentiality reasons, as it obstructs secondary studies.

| Viewpoint | Consideration | Recommendation |
|---|---|---|
| Tool developer | Which IR model should we implement in our new trace recovery tool? | The classic VSM, since several efficient implementations are available as open source. There is no empirical evidence that more advanced IR models produce more accurate trace links. |
| Development manager | Should we deploy an IR-based trace recovery tool in our organization? | Await empirical evidence from future in-vivo studies. In the meantime, assure that your general search solutions make trace artifacts findable. |
| Traceability researcher | How can we strengthen the base of empirical evidence of IR-based trace recovery? | Case studies in industrial settings are required. Furthermore, larger datasets containing industrial artifacts should be used as experimental input. Also, the reporting of evaluation contexts, input artifacts, and IR solutions should be improved. |

**Table 3:** Recommendations for the proposed viewpoints.

## RQ5: How can we advance human-oriented studies on IR-based trace recovery?

As paper I shows, a majority of evaluations of IR-based trace recovery have been conducted in "the cave of IR evaluation", drawing conclusions based on P-R curves. More evaluations with human subjects, working with realistic tasks, are needed to strengthen the evidence of IR-based trace recovery. While a number of controlled experiments have been conducted, conspicuously few industrial case studies have been reported. In an attempt to guide future studies beyond "the cave", Paper III proposes a taxonomy of evaluation contexts, along with suggested measures, tailored for IR-based trace recovery, based on previous work by Ingwersen and Järvelin [51].

In Table 3, we further summarize our answers in an attempt to address the three viewpoints presented in Section 3. The last column presents our recommendations, based on the understanding obtained during the work of this thesis.

## 7 Threats to Validity

The results of any research effort should be questioned, even though proper research methodologies were applied. The validity of the research is the foundation on which the trustworthiness of the results is established. In this thesis, threats to validity, and actions taken to reduce the threats, are discussed based on the classification proposed by Wohlin *et al.* [95]. Further details on validity threats are available in the individual papers.

*Construct validity* is concerned with the relation between theories behind the research and the observations. Consequently, it covers the choice and collection of measures for the studied concepts. For example, the questions of a questionnaire must not be misunderstood or misinterpreted by the respondents. One strategy to increase construct validity is to use multiple sources of evidence, and to establish chains of evidence [96].

In Paper I, we partly aggregate evidence from previous evaluations based on data in tables or directly from P-R curves. Thus, we were limited by the levels of detail included in the reviewed publications. A possible way to obtain richer data would have been to contact the corresponding authors and ask for access to all measurements from the studies. As 79 publications from the last decade were included, it would have required a large effort. On the other hand, as we extracted data from P-R values from 48 publications and, whenever possible, followed the TREC convention of reporting both precision at fixed recall levels as well as precision and recall at certain cut off levels, we limit this threat. Regarding the survey in Paper II, the questionnaire was reviewed by a native English speaker, and a pilot study was conducted on five senior software engineering researchers.

*Internal validity* is related to issues that may affect the causal relationship between treatment and outcome. In experiments, used in both Paper III and IV, the internal validity questions whether the effect is caused by the independent variables or other factors. Internal validity is typically not a threat to descriptive or exploratory studies, as casual claims rarely are made [96].

The SLR in Paper I is subject to a number of threats to internal validity. As most evaluations of IR-based trace recovery have been conducted in controlled settings, e.g., in university classrooms, we have not considered different domains in the analysis of the results. Further research is required to study whether the approach is more feasible in certain contexts such as safety-critical development. Also, as the use of terminology in the publications was not aligned, our choice of search string might have influenced the resulting evidence base. These threats were addressed by combining database searches with snowball sampling, and by incrementally developing the search string based on a gold standard of publications. Another threat to the SLR is publication bias, e.g., authors might be less likely to publish negative results, or IR-based trace recovery might be successfully used in industry even though it is not reported in research publications. As the results in Paper I are related to all RQs, so are the threats to internal validity. Furthermore, regarding the experiments included in this thesis, our understanding of the studied IR-based trace recovery tools (Paper III), and the subjects' understanding of the work task (Paper IV), are confounding factors. In both experiments we addressed threats to internal validity by running pilot experiments.

*External validity* concerns the ability to generalize the findings outside the actual scope of the study. Results obtained in a specific context may not be valid in other contexts. Strategies to address threats to external validity include studying multiple cases and replicating experiments [96].

In Paper II, we surveyed published researchers in the traceability community. However, as the number of respondents was low, we do not have a strong basis for generalizing our results to the entire population of traceability researchers. On the other hand, considering the exploratory nature of our study, the external validity of the survey is acceptable. Another threat to external validity is that all software artifacts used as experimental input in Paper III and IV originate from embedded development contexts, either from the space domain or process automation. Furthermore, as emphasized in Paper I, the limited number of artifacts makes generalizations to larger document spaces uncertain.

*Conclusion validity* results from the ability to draw correct conclusions about the relation between the treatment and the outcome. This type of validity is related to the repeatability of a study. Threats to conclusion validity in quantitative studies are often related to statistics [95]. In qualitative studies on the other hand, it can be used to discuss to which extent the data and the analysis are dependent on the specific researchers, then sometimes also referred to as reliability [84, 96].

In Paper III, the technology-oriented experimental results were not analyzed using significance testing since assumptions underlying statistical treatment such as independence, random sampling and normality were not met. Instead, the output differences were analyzed in a human-oriented experiment in Paper IV. While these results were analyzed using statistical testing, the low number of subjects did not result in any statistically significant results. On the other hand, we consider the effect sizes reported in Paper IV to be of practical significance.

In Paper I, we assess the strength of evidence of the industrial feasibility of IR-based trace recovery. This assessment involves interpretation. While we do not consider P-R curves from small evaluations "in the cave" to be particularly strong pieces of evidence, other researchers might value them differently. For example, the foreword by Finkelstein in the recently published textbook on software and systems traceability discusses IR-based recovery in a less critical manner [18]. However, in line with practices in reflexive methodology, there is a demand for reflection in research in conjunction with interpretation [3]. As such, a researcher should be aware of, and critically confront, favored lines of interpretation. Naturally, there is a risk of bias in the foreword of a textbook, written by a notable part of the traceability research community. In such a foreword, there are few incentives to present sceptical views on the research. On the other hand, there is also a risk that the work in this licentiate thesis, written by a junior PhD student fostered in a strictly empirical research tradition, is overly critical to the mainly technology-oriented research strategy. Our conclusion, that there is a need for evaluations that go beyond "the cave", might be in the interest of the individual researcher, as an attempt to pave the way for future empirical studies. To conclude, the conclusion validity is a threat to RQ5 and RQ6, as other researchers might suggest different ways to advance evaluation of IR-based trace recovery.

## 8   Agenda for Future Research

This section presents a speculative research agenda for future work, partly based on Paper V. We intend to continue our research with a focus on trace links, however in a more solution-oriented manner. Our ambition is to study a specific work task that requires an engineer to explicitly specify trace links among artifacts, namely change impact analysis in a safety-critical context. As we suspect that software engineers are more comfortable navigating the source code than its related documentation, we intend to focus specifically on trace links between non-code artifacts. A summary of the planned work in this section is presented as Future research Questions (FQ) and planned Design science Tasks (DT) in Table 5.

### 8.1   Description of the Context

The targeted impact analysis process is applied by a large multinational company active in the power and automation sector. The development context is safety-critical embedded development in the domain of industrial control systems, governed by IEC 61511 [52]. The number of developers is in the magnitude of hundreds; a project has typically a length of 12-18 months and follows an iterative stage-gate project management model. Also, the software is certified to a Safety Integrity Level (SIL) of 2 as defined by IEC 61508 [53], corresponding to a risk reduction factor of 1.000.000-10.000.000 for continuous operation. Process requirements mandate maintenance of traceability information, especially between requirements and test cases. Both requirements and test case descriptions are predominantly specified in English NL text.

As specified in IEC 61511 [52], impact of proposed software changes, e.g., for error corrections, should be analyzed before implementation. In the initially studied case, as presented in Paper V, this process is integrated in the issue tracking system. As part of the analysis, engineers are required to investigate impact, and report their results according to a project specific template, validated by an external certifying agency. A slightly modified version of this template, recently described as part of a master thesis project [64], is presented in Table 4. As seen in Table 4, several questions explicitly ask for trace links (6 out of 13 questions). The engineer is required to specify source code that will be modified (with a file-level granularity), and also which related software artifacts need to be updated to reflect the changes, e.g., requirement specifications, design documentation, test case descriptions, test scripts and user manuals. Furthermore, the impact analysis should specify which high-level system requirements cover the involved features, and which test cases should be executed to verify that the changes are correct once implemented in the system. Consequently, the impact analysis reports explicitly connect requirements and test artifacts. As this has been reported as a specific challenge in requirements and verification alignment [85], we also intend to explore how the knowledge embedded in the impact analysis reports can be used to support this aspect of large-scale software development.

| | Impact Analysis Questions for Error Corrections |
|---|---|
| 1) | Is the reported problem safety critical? |
| 2) | In which versions/revisions does this problem exist? |
| 3) | How are general system functions and properties affected by the change? |
| 4) | **List modified code files/modules and their SIL classifications, and/or affected safety safety related hardware modules.** |
| 5) | How are general system functions and properties affected by the change? |
| 6) | **Which library items are affected by the change? (e.g., library types, firmware functions, HW types, HW libraries)** |
| 7) | **Which documents need to be modified? (e.g., product requirements specifications, architecture, functional requirements specifications, design descriptions, schematics, functional test descriptions, design test descriptions)** |
| 8) | **Which test cases need to be executed? (e.g., design tests, functional tests, sequence tests, environmental/EMC tests, FPGA simulations)** |
| 9) | **Which user documents, including online help, need to be modified?** |
| 10) | How long will it take to correct the problem, and verify the correction? |
| 11) | What is the root cause of this problem? |
| 12) | How could this problem been avoided? |
| 13) | **Which requirements and functions need to be retested by product test/system test organization?** |

**Table 4:** Impact analysis template. Questions in bold fonts require explicit trace links to other artifacts. Based on a description by Klevin [64].

## 8.2 Solution idea

While an important part of the impact analysis work task involves specifying trace links to related software artifacts, there are rarely any traceability matrices to consult. Consequently, if engineers do not already know which artifacts are impacted, a substantial part of the impact analysis work task turns into an information seeking activity. In Figure 5, we present an initial model of the trace link seeking activity involved in the impact analysis. At first, depicted in the left of the figure, the engineer starts the work task with six questions that require explicit trace links. The engineer then enters the process of trace link seeking, presented as the second step in Figure 5. Typically, this is an iterative process where the engineer seeks information suggesting trace links in different ways. Knowledge embedded in previous impact analysis reports can be reused, project documentation can be studied, and colleagues can be asked. As reported by Dagenais *et al.*, especially junior engineers and newcomers rely on communication with more experienced colleagues, in particular when project findability is low due to poor search solutions [21]. Finally, as presented to the right in Figure 5, enough information has been found to specify required trace links in the impact analysis template. As presented in Table 5, we intend to improve the trace link seeking model (DT1) based on observational studies with protocol analysis. This work could complement Freund *et al.*'s more general work on modeling the information behavior of software engineer [37] by exploring a specific work task. Moreover, we plan to assess whether the trace link seeking model is applicable to other contexts with strict process requirements on maintenance of traceability information (FQ1).

Currently, as presented in Paper V, engineers conduct the trace link seeking supported by a low level of automation [75]. Our plan is to increase the level of automation in two areas of the trace link seeking process, as indicated by the cogwheels in Figure 5. In the present work flow, engineers use the search features (primarily keyword-based) of the issue tracking system and the document management system to gather enough information to specify trace links. Our hypothesis is that these steps could be supported by a recommendation system based on textual similarity analysis. As discussed in Paper V, our goal is to support trace link seeking by deploying a plug-in to the issue tracking system (presented as DT2 in Table 5). Developing plug-ins to tools already deployed in industry enables in-vivo studies without introducing additional external tools.

Another direction we want to explore is to consider artifact meta-information to improve the trace recovery, presented as FQ2 in Table 5. One possibility, that we initially have explored, is to exploit the already existing link structures among software artifacts. Using link mining, we have explored clusters of issue reports from the public Android issue tracking system. Figure 6 visualizes link structures among Android issue reports, extracted from hyperlinks manually established by developers. We expect to find patterns of linked artifacts also in the targeted safety-critical case, however also between different types of artifacts, when conducting link mining in the impact analysis reports in the issue tracking system. As hyperlinks have proven useful in tasks such as object ranking, link prediction, and subgraph discovery [38], we hope it can also be used to advance trace recovery. A link mining approach might move our research closer to work on semantic networks of software artifacts, which previously has been used to significantly improve search-ing based on textual similarity in the software engineering context [58]. Furthermore, work on trace link structures would enable us to explore the use of visualization techniques to support engineers' trace links seeking, as has previously been proposed by Cleland-Huang and Habrat [19].

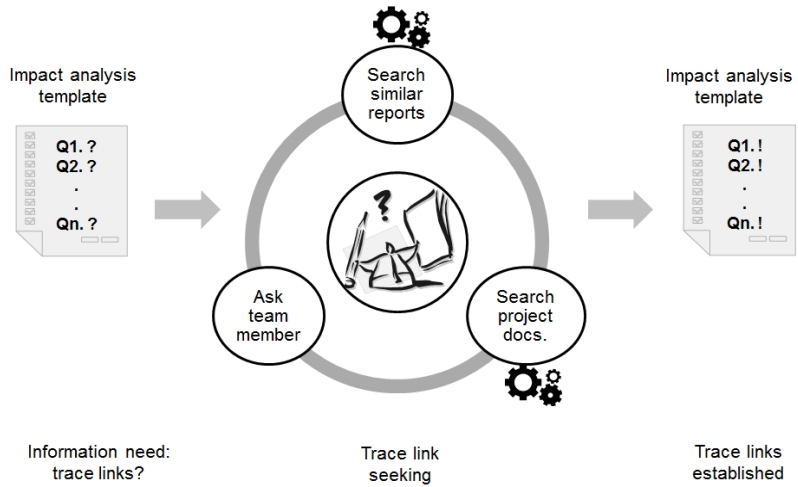

**Figure 5:** Trace link seeking in the impact analysis work task. Adapted from Paper V. Cogwheels indicate an information seeking activity that could be supported by IR-based trace recovery.

We also suspect that other pieces of artifact meta-information could be useful in trace recovery. Web search engines consider hundreds of features to assess the relevance of web pages for ranking purposes [1]. Learning-to-rank methods are then used on training data to learn the optimal combination of feature weights, resulting in the best ranking of search results [67]. In the context of trace recovery, we envision that both nominal software artifact features (e.g., responsible team, subsystem), ordinal features (e.g., safety level, severity), and features measurable on a ratio scale (e.g, resolution time, link structure) can be used to improve ranking of candidate trace links, in particular when combined with information about the user of the tool. Engineers conducting trace recovery might not consider the relevance of candidate trace links to be binary, but rather of a multi-dimensional nature [61], i.e., dynamic and situational. For example, the relevance of a trace link might depend on the role of the tracing engineer (tester, developer, manager, etc.), the current phase of the development project (pre-study, implementation, verification, etc.), and which other trace links have already been identified (as there might be dependencies). Using meta-information and user information, IR-based trace recovery could assumably be advanced beyond what is possible using merely textual similarity analysis.

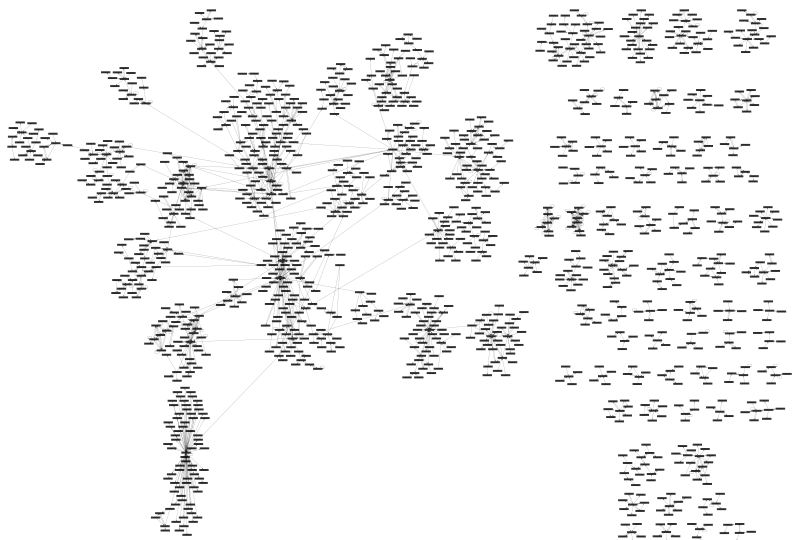

**Figure 6:** Linked structures of issue reports in the public Android issue tracking system.

We anticipate certain challenges as we continue our work. First, in many enterprises, information access is hindered by information being widely dispersed in information management systems with poor interoperability [69], resulting in what is referred to as *information silos*. It is uncertain which artifacts could be accessed without major engineering efforts and without breaking information access policies. Second, as identified by Klevin [64], the impact analysis reports in the targeted case, i.e., the answers to the template presented in Table 4, are stored in the issue tracking system as unstructured text. Clearly, this will complicate information extraction and data mining from the reports. Third, while the number of software artifacts in large projects can be challenging, it is several orders of magnitude smaller than the number of web pages indexed by modern web search engines. There is a risk that we will not be able to gather enough data for machine learning methods to do themselves justice.

## 9 Conclusion

The challenge of maintaining trace links in large-scale software engineering has been addressed by IR approaches in roughly a hundred previous publications. In an SLR, we identified 79 publications reporting empirical evaluations of IR-based trace recovery. We found that most often algebraic IR models have been applied (RQ1), and confirm the previous claim that no IR model regularly outperforms trace recovery based on VSM (RQ2).

| Future work | Description | Research method | Type of research |
|---|---|---|---|
| DT1 | How can we further improve the trace link seeking model? | Design science | Modeling |
| FQ1 | Is the trace link seeking model applicable in other development contexts with process requirements on traceability? | Multi-case study | Exploratory |
| DT2 | How can textual similarity analysis be applied to support trace recovery in the impact analysis? | Design science | Tool development |
| FQ2 | Can the accuracy of the tool output be improved by considering artifact meta-information? | Technology-oriented experiment | Evaluative |
| FQ3 | Does the tool support the impact analysis work task? | Case study | Evaluative |

**Table 5:** Future research questions and planned design science tasks.

A majority of previous evaluations of IR-based trace recovery have been technology-oriented, conducted in what Ingwersen and Järvelin refer to as "the cave of IR evaluation". Also, we show that evaluations of IR-based trace recovery primarily have been conducted using simplified datasets, both in relation to size (most often less than 500 artifacts) and origin (frequently former student projects, typically not validated for industrial representability). As such, the validity of concluding that IR-based trace recovery is feasible in an industrial setting, based on P-R curves "in the cave", can be questioned (RQ3).

On the other hand, a set of previous evaluations conducted with human subjects suggest that engineers would benefit from IR-based trace recovery tools when performing certain work tasks (RQ3). To further strengthen the evidence of IR-based trace recovery, more studies involving humans are needed, particularly industrial case studies (RQ5). Moreover, evaluative studies should be conducted on diverse datasets containing a higher number of artifacts (RQ4). Consequently, our findings intensify the call for additional empirical research by CoEST.

This thesis also includes two experiments on IR-based trace recovery. The technology-oriented experiment highlights the clear dependence between datasets and the accuracy of IR-based trace recovery, which was also confirmed by the SLR. Thus, to enable replications and secondary studies, we argue that datasets should be thoroughly characterized in future studies on trace recovery, especially when they cannot be disclosed (RQ4). The human-oriented experiment suggests that it is worthwhile investigating further into the actual value of improved P-R curves. The pilot experiment showed that the effect size of using a slightly better tool is of practical significance regarding precision and F-measure. Finally, based on research on general IR evaluation, we propose a taxonomy of evaluation contexts tailored for IR-based trace recovery (RQ5).

As future work, we intend to target an industrial case of impact analysis in a safety-critical development context. In the case, engineers perform trace link seeking among textual artifacts, and explicitly specify the trace links according to a template. Consequently, the case appears to be suitable for evaluating IR-based trace recovery in an in-vivo setting.